\documentclass[aps,prd,floatfix,showpacs,10pt,nofootinbib]{revtex4-1}

\usepackage{amsmath,amssymb,amsfonts,amsthm} %usepackage{amsthm,amsfonts,amssymb,amsmath,amsbsy}
\usepackage{color}
\usepackage{float}
\usepackage{graphicx}
\usepackage{pstricks,pstricks-add}
\usepackage{pst-plot}
\usepackage[scriptsize,nooneline,hang]{caption}
\usepackage[hang,nooneline,scriptsize]{subfigure}

\definecolor{dark-green}{rgb}{0,0.7,0}
\definecolor{dark-blue}{rgb}{0,0.2,0.5}
\definecolor{med-blue}{rgb}{0,0.7,1}
\definecolor{mblue}{rgb}{0,0.2,1}
\definecolor{cnc}{rgb}{0.8,0,0}
\definecolor{light-red}{rgb}{1,0.8,0.8}
\definecolor{dark-yellow}{rgb}{1,0.8,0}
\definecolor{light-blue}{rgb}{0.8,0.9,1}
\definecolor{grey}{rgb}{0.211,0.211,0.211}
\definecolor{verylight-blue}{rgb}{0.93,0.95,1}
\definecolor{light-yellow}{rgb}{1,0.9,0.8}
\newcommand{\weglassen}[1]{}

\begin{document}

\title{Abelian-Higgs strings in Rastall gravity}

\author{Eug\^enio R. Bezerra de Mello $^{(a)}$ }
\email{emello@fisica.ufpb.br    }

\author{J\'ulio C. Fabris $^{(b)}$}
\email{fabris@pq.cnpq.br   }

\author{Betti Hartmann $^{(b), (c)}$ }
\email{bettihartmann@googlemail.com}
%\affiliation{School of Engineering and Science, Jacobs University Bremen, 28759 Bremen, Germany}

\affiliation{
$(a)$ Departamento de F\'{\i}sica, Universidade Federal da Para\'{\i}ba, 58.059-970, Caixa Postal 5.008, Jo\~{a}o Pessoa, PB, Brazil\\
$(b)$ Universidade Federal do Esp\'irito Santo, Departamento de F\'isica, CEP 29075-910, Vit\'oria (ES), Brazil\\
$(c)$ School of Engineering and Science, Jacobs University Bremen, 28759 Bremen, Germany}

%\maketitle

\date\today

\begin{abstract}
In this paper we analyze Abelian-Higgs strings in a phenomenological model that takes quantum effects in curved space-time into account.
This model, first introduced by Rastall, cannot be derived from an action principle. We formulate phenomenological equations of
motion under the guiding principle of minimal possible deformation of the standard equations. We construct string solutions
that asymptote to a flat space-time with a deficit angle by solving the set of coupled
non-linear ordinary differential equations numerically. Decreasing the Rastall parameter
from its Einstein gravity value we find that the deficit angle of the space-time increases and becomes
equal to $2\pi$ at some critical value of this parameter that depends on the remaining couplings in the model.
For smaller values the resulting solutions are {\it supermassive} string solutions possessing a singularity at a finite
distance from the string core. Assuming the Higgs boson mass to be on the order of the gauge boson mass
we find that also in Rastall gravity this happens only when the symmetry breaking scale is on the order of the Planck mass. 
We also observe that for specific values of the parameters in the model the energy per unit length becomes
proportional to the winding number, i.e. the degree of the map $S^1\rightarrow S^1$. Unlike in the BPS limit in Einstein gravity, this is, 
however, not connect to an underlying mathematical
structure, but rather constitutes a {\it would-be-BPS bound}.
\end{abstract}


\maketitle

\section{Introduction}
\label{Int}

One of the well-known ingredients of Einstein's theory of General Relativity is the covariant conservation of the energy-momentum tensor
which leads, via Noether's Theorem, to the conservation of globally defined quantities. These quantities appear as integrals of the
components of the energy-momentum tensor over suitable space-like surfaces that typically have one of the Killing vectors
of the space-time as their normal. As such, the total rest energy/mass of a system is conserved in
General Relativity. Now the question is whether this is a suitable assumption as there is (up to date) no clear experimental evidence for this.
Hence, models have been developed that relax the condition of covariant energy-momentum conservation. In this paper we are interested in a modification
of General Relativity suggested by Rastall \cite{Rastall:1973nw}. In this model, the gravitational fields are sourced by the energy and momentum, as in 
General Relativity, but also by the metric of the external space. Since in empty space, i.e. for a vanishing energy-momentum tensor Rastall
gravity agrees with General Relativity, this can be seen as a direct implementation of Mach's principle stating that the inertia of a mass distribution
should be dependent on the mass and energy of the external space-time \cite{Majernik:2006jg}.
This model has been studied extensively in the context of cosmology \cite{Batista:2012hv,Fabris:2012hw,Batista:2011nu}.

At first sight Rastall's theory has major drawbacks: its phenomenological formulation and in addition the absence of a 
variational principle. However, it contains a rich structure that may be easily connected with many fundamental aspects of a 
gravity theory. First of all, the usual energy-momentum conservation law of Special Relativity may be generalized to curved 
space-time in many different ways, including geometric terms. General Relativity is one possible extension and constitutes a minimal 
implementation of such a generalization by replacing the standard derivative with a covariant derivative.
This by itself is not completely free of unclear aspects. On the other hand, if quantum effects are taken into account in curved space-time
the classical expression for the energy-momentum tensor must be modified introducing quantities related to the 
curvature of the space-time \cite{bd}. Moreover, the propagation of quantum fields in space-times with horizons
may lead to a violation of the classical conservation law (due to the chirality of the quantum modes)
leading to a so-called {\it gravitational anomaly} \cite{anomalias}.
In this sense, Rastall's theory is a phenomenological procedure to consider effects of quantum fields in curved space-time
and to investigate in a completely covariant way such effects. Even if there is in principle no action leading to the 
Rastall equation it is possible to find such an action
if an external field is introduced in the Einstein-Hilbert action through a Lagrange multiplier. 
This is somehow a reminiscence of the quantum effects described by the Rastall equation. 
Other geometrical frameworks, like Weyl geometry, may lead to equations similar to the ones in Rastall gravity \cite{smalley,romero}.

In this paper we are interested in static, cylindrically symmetric solutions to the Rastall gravity model coupled to 
the U(1) Abelian-Higgs model. These solutions constitute field theoretical realizations of a specific type of 
topological defect called Abelian-Higgs string \cite{no} which could e.g. describe infinite straight cosmic strings
whose properties have been analyzed in \cite{DG,Laguna}.
Topological defects are believed to have formed in the numerous phase transitions in the early
universe due to the Kibble mechanism \cite{topological_defects}.
While magnetic monopoles and domain walls, which result from the spontaneous
symmetry breaking of a spherical and parity symmetry, respectively, 
are catastrophic for the universe since they would overclose it, cosmic strings
are an acceptable remnant from the early universe. These objects 
form whenever an axial symmetry gets spontaneously broken and, due to topological arguments,
are either infinitely long or exist in the form of cosmic string loops. Numerical
simulations of the evolution of cosmic string networks have shown that
these networks reach a scaling solution, i.e. their contribution to the total energy density
of the universe becomes constant at some stage. The main mechanism that allows
cosmic string networks to reach this scaling solution is the formation
of cosmic string loops due to self-intersection and the consequent decay of these loops
under the emission of gravitational radiation.

For some time, cosmic strings were believed to be responsible for the structure
formation in the universe. New Cosmic Microwave Background (CMB) data clearly
shows that the theoretical power spectrum associated to cosmic strings
is in stark contrast to the observed power spectrum. However, other effects might be
caused by moving cosmic strings that can potentially be observed in the CMB data (see e.g. \cite{Planck} for a recent discussion). 
Moreover, there has been
a recent revival of cosmic strings since it is now believed that cosmic strings
might be linked to the fundamental strings of String Theory \cite{polchinski}.
While perturbative fundamental strings were excluded to be observable on cosmic scales
for many reasons \cite{witten}, there are now new theories containing
extra dimensions, so-called brane world model, that allow to lower the fundamental
Planck scale down to the TeV scale. 
Moreover, cosmic strings are interesting due to the recent BICEP2 data \cite{bicep} that showed evidence
for a B-mode polarization at small ${\ell}$ in the Cosmic Microwave background (CMB). While topological defects cannot be accounted for the
B-mode polarization alone \cite{Lizarraga:2014eaa} it has been suggested that this polarization results from
gravitational waves, i.e. tensor modes, originating from an inflationary epoch in the early universe. If that turns out to be
correct the question remains what the origin of the field driving inflation is and how it can be embedded into
suitable Unified Theories, which are certainly necessary to describe the physics at the energy scales relevant for the inflationary epoch. 
Now, it is interesting that
cosmic strings generically form at the end of inflation in inflationary models
resulting from String Theory \cite{braneinflation} and Supersymmetric Grand Unified Theories \cite{susyguts}.
Hence, it is conceivable that cosmic strings show up in the CMB and, indeed, CMB data (power and polarization spectra)
are well compatible with a substantial amount of the total energy density of the universe coming from cosmic strings \cite{cmb,cmb2}.

While field theoretical cosmic strings would always form loops when self-intersecting (and hence providing a ``short-cut'' for the
magnetic flux), this can well be different for cosmic superstrings. The question then is how these networks of cosmic superstrings
loose energy to reach a scaling solution and hence be not dangerous for the universe today. 
One of the suggestions is that they can form bound states. Since it is difficult to
study cosmic superstrings with respect to the formation of bound states, the interaction of cosmic 
strings has been investigated in the context of field theoretical
models describing bound systems of D- and F-strings, so-called p-q-strings \cite{saffin,hu}. 

In order to understand the lensing properties of cosmic strings (and hence their impact on the CMB spectrum) as well as the 
evolution of cosmic string networks in different gravity models it is important to study their properties in detail.
This is the aim of this paper which considers cosmic string in the Rastall gravity model. 

Our paper is organized as follows: In Section \ref{Model}, we briefly introduce the gravity model proposed by Rastall. 
In Section \ref{Cosmic} we present the U(1) Abelian-Higgs model that we study coupled to Rastall gravity. 
We present the set of differential equations associated with this system and give the asymptotic
behavior for the matter and for the metric functions. In Section \ref{Numerical} we present our numerical results.
In section \ref{Conc} we give our conclusions. The Appendix contains an outline
on the procedure we used to derive the equations of motion for the matter fields.

\section{The model}
\label{Model}

Rastall's generalization of General Relativity \cite{Rastall:1973nw} uses the idea of covariant non-conservation of the energy-momentum tensor and has been
cast into the following form (where $\mu, \nu=0,1,2,3$)
\begin{equation}
\label{eq1}
 D_{\mu} T^{\mu\nu} = \kappa D^{\nu} R  \   ,
\end{equation}
or 
\begin{equation}
\label{eq2}
 D_{\mu} T^{\mu\nu} = \bar{\kappa} D^{\nu} T  \  , 
\end{equation}
where $D_{..}$ denotes the covariant derivative, 
$T^{\mu\nu}$ the energy-momentum tensor with trace $T\equiv T^{\mu}_{\mu}$ and
$R$ the Ricci scalar. $\kappa$ and $\bar{\kappa}$ are some coupling constants such that in the limit $\kappa\rightarrow 0$ 
we recover standard Einstein gravity. Note that the equations above imply a violation of the principle
of General Covariance. 

The equations for Rastall gravity and the violation of the conservation of energy-momentum tensor then read 
\begin{equation}
\label{e1}
 R_{\mu\nu}-\frac{1}{2} g_{\mu\nu} R = 8\pi G \left( T_{\mu\nu} - \frac{\gamma-1}{2} g_{\mu\nu} T\right) 
\end{equation}
and
\begin{equation}
\label{e2}
D_{\mu} T^{\mu\nu}=\frac{\gamma-1}{2} D^{\nu} T  \ .
\end{equation}
Of course the constants $\kappa$, $\bar\kappa$ and $\gamma$ can be easily related and $\gamma=1$ corresponds to the
Einstein gravity limit.

The equations \eqref{e1} and \eqref{e2} can be written in the form
\begin{equation}
\label{mod_gravity}
 R_{\mu\nu}=8\pi G \left(T_{\mu\nu} - \frac{1}{2}g_{\mu\nu} T + \frac{1}{2} g_{\mu\nu} (\gamma-1)T\right) 
\end{equation}
and
\begin{equation}
\label{mod_conservation}
 \frac{1}{\sqrt{-g}} \partial_{\mu}\left(\sqrt{-g} T^{\mu\nu}\right) + \Gamma^{\nu}_{\mu\lambda} T^{\mu\lambda} 
= \frac{\gamma-1}{2} \partial^{\nu} T  \  , 
\end{equation}
where $\Gamma^{\nu}_{\mu\lambda}$ are the Christoffel symbols. 

\section{Abelian-Higgs strings}
\label{Cosmic}

In this section  we would like to study Abelian-Higgs strings in Rastall gravity. 
The matter Lagrangian density, $\mathcal{L}_{\rm m}$,  is given by 
\begin{eqnarray}
\mathcal{L}_{\rm m}&=&D_{\mu}\phi(D^{\mu}\phi)^*-
\frac{1}{4}F_{\mu\nu}F^{\mu\nu}-\frac{\lambda}{2}(\phi\phi^*-\eta^{2})^2 \  ,
\end{eqnarray}
with the covariant derivative  
$D_{\mu}\phi$ = $\nabla_{\mu}\phi$ - $i e A_{\mu}\phi$
of the complex scalar field
$\phi$.  The field strength tensor is 
$F_{\mu\nu}=\nabla_{\mu}A_{\nu}-\nabla_{\nu}A_{\mu}=\partial_{\mu}A_{\nu} - \partial_{\nu}A_{\mu}$, 
of the U(1) gauge potential $A_{\mu}$ with coupling constant $e$. 
$\nabla_{\mu}$ denotes the gravitational covariant derivative. 
Finally, $\lambda$ is the self-coupling of the scalar field, while 
$\eta$ denotes the vacuum expectation value. We define the 
energy momentum tensor to be given in the standard way by the variation of the matter Lagrangian ${\cal L}_{\rm m}$ 
with respect to the metric
\begin{equation}
 T_{\mu\nu} = - 2 \frac{\delta {\cal L}_{\rm m}}{\delta g^{\mu\nu}} + g_{\mu\nu} {\cal L}_{\rm m}  \  .
\end{equation}
Note, however, that this is not linked to an action principle in which the variation of the total action (matter plus metric) with respect
to the metric gives the gravity equations. The model we are studying here is a phenomenological model in which we insert ``by hand'' the 
non-conservation of the energy-momentum tensor. 

The symmetry breaking pattern is such that $U(1) \rightarrow 1$ and the scalar field as well as the gauge field
acquire mass: the Higgs field  has mass $M_{H}=\sqrt{2\lambda} \eta$, 
while the gauge boson mass is $M_{W}=\sqrt{2}e \eta$.

By using the standard cylindrical coordinates $(r,\varphi,z)$, the most general 
static cylindrically symmetric line element invariant under boosts along the $z$-direction is:
\begin{eqnarray}
\label{metric}
ds^2=N^2dt^2-dr^2-L^2d\varphi^2-N^2dz^2  \  , \label{cysymmetric}
\end{eqnarray}
where $N$ and $L$ are functions of $r$ only.

The non-vanishing components of the Ricci tensor $R_{\mu}^{\nu}$ then read \cite{clv}:
\begin{eqnarray}
R_t^t=-\frac{(LNN')'}{N^2 L} \ \ , \ \ R_{r}^{r} = -\frac{2N''}{N}-\frac{L''}{L} \ \ \ , \ \ 
R_{\varphi}^{\varphi}= -\frac{(N^2 L')'}{N^2 L} \ \ \ , \ \ \ R_z^z=R_t^t
\end{eqnarray}
where the prime now and in the following denotes the derivative with respect to $r$. 

For the matter and gauge fields, we have \cite{no}
\begin{equation}
\phi(r,\varphi)=\eta f(r)e^{i n\varphi} \ \ , \ \  
A_{\mu}dx^{\mu}=\frac {1}{e}(n-P(r)) d\varphi \ \ , 
\end{equation}
where $n$ is an integer indexing the vorticity of the Higgs field  around the $z-$axis, i.e. corresponds 
to the degree of the map $S^1 \rightarrow S^1$. In the following we will study only the case
$n=1$. 

Now let us define the following dimensionless quantities
\begin{equation}
\label{scaling}
r\rightarrow \frac{r}{e\eta} \ \ \ , \ \ \ L\rightarrow \frac{L}{e\eta} \  ,
\end{equation}
such that $r$ measures the radial
distance in units of $M_{W}/\sqrt{2}$. 
Then, the Lagrangian density, ${\cal L}_m \rightarrow {\cal L}_m/(\eta^4 e^2)$, depends only on 
the following dimensionless coupling constants
\begin{equation}
\alpha=8\pi G\eta^2 = 8\pi \frac{\eta^2}{M_{\rm pl}^2}  \ \ ,  \ \ 
\beta=\frac{\lambda}{e^2}=\frac{M^2_{H}}{M^2_{W}} \  ,
\end{equation}
where $M_{\rm pl}$ is the Planck mass.

The non-vanishing components of the energy-momentum tensor read \cite{clv}
\begin{equation}
 -T^t_t = (f')^2 +  \frac{(P')^2}{2L^2} + \frac{P^2 f^2}{L^2}+
\frac{\beta}{2}\left(f^2-1\right)^2   \  \ , \ \ T^z_z=T^t_t \ \ , \ \ 
 \end{equation}
\begin{equation}
 -T^r_r=-(f')^2 - \frac{(P')^2}{2L^2} + \frac{P^2 f^2}{L^2}+
\frac{\beta}{2}\left(f^2-1\right)^2 \ \ ,
\end{equation}
\begin{equation}
 -T^{\varphi}_{\varphi}=(f')^2 - \frac{(P')^2}{2L^2} - \frac{P^2 f^2}{L^2}+
\frac{\beta}{2}\left(f^2-1\right)^2 \ \ 
\end{equation}
and the trace $T\equiv T^{\mu}_{\mu}$ is given by
\begin{equation}
 T= -2\left[(f')^2  + \frac{P^2 f^2}{L^2}+
\beta\left(f^2-1\right)^2 \right]  \  . 
\end{equation}

Now, since the trace of the energy-momentum tensor is independent of the $t$ and the $z$ coordinate we have
\begin{equation}
\label{conserved}
 D_{\mu} T^{\mu t}= 0 \ \ , \ \    D_{\mu} T^{\mu z}=0  \ . 
\end{equation}
Note that this is not a consequence of the gravity model used, as it would be in General Relativity, but rather a consequence of the
particular choice of the matter content. 
Then, (\ref{conserved}) allows us to define globally conserved charges, namely the energy per unit length, $\mu$,  as well as the 
tension along the string axis, $\tau$,  in the usual way (with $^{(2)} g$ denoting the determinant of the induced metric
on spatial sections perpendicular to $z$)
\begin{equation}
\mu=\int_0^{2\pi}\int_0^\infty \sqrt{^{(2)}g} \  T^t_t \ dr d\theta= 2\pi  \int_0^\infty \  L \ T^t_t \ dr \ \  {\rm and}  \  \
\tau=\int_0^{2\pi}\int_0^\infty \sqrt{^{(2)}g} \ T^z_z \ dr  d\theta =  2\pi  \int_0^\infty \  L \ T^z_z \ dr  \  ,
\end{equation}
where obviously from the fact that $T^t_t=T^z_z$ we find that $\mu=\tau$. Note that we ``measure'' $\mu$ in units of $\eta^2$. 
A quantity often cited when discussing the observational effects of cosmic strings is $G\tilde{\mu}$, where $\tilde{\mu}$ is the
dimensionful energy per unit length of the string. This quantity enters in both the expression for the deficit angle
as well as the temperature anisotropies $\Delta T/T$ in the CMB. In our dimensionless units the
quantity $G{\tilde{\mu}}$ is equal to $\alpha \mu/(8\pi)$.

\subsection{Equations of motion}

We use the $(tt)$- and $(\varphi\varphi)$-components of the Rastall equation (\ref{mod_gravity}):
\begin{eqnarray}
\label{eq_grav1}
\frac{(LNN')'}{N^2 L}&=& \alpha\left[(\gamma-1) (f')^2 + \frac{(P')^2}
{2 L^2}+ (\gamma-1) \frac{P^2 f^2}{L^2} + \left(\gamma-\frac{3}{2}\right)\beta(f^2-1)^2\right]
\end{eqnarray}
and 
\begin{eqnarray}
\label{eq_grav2}
\frac{(N^2L')'}{N^2L}&=&\alpha\left[(\gamma-1)(f')^2 - \frac{(P')^2}{2 L^2}+
(\gamma-3)\frac{P^2 f^2}{L^2} + \left(\gamma-\frac{3}{2}\right) \beta (f^2-1)^2 \right]  \  .
\end{eqnarray}
The $(rr)$ component reads
\begin{equation}
 \frac{2N''}{N}+\frac{L''}{L}=
 \alpha\left[(\gamma-3)(f')^2 - \frac{(P')^2}{2L^2} + (\gamma-1)\frac{P^2 f^2}{L^2}+\left(\gamma-\frac{3}{2}\right)\beta(f^2-1)^2\right] \  ,
\end{equation}
which can be combined with (\ref{eq_grav1}) and (\ref{eq_grav2}) to obtain a constraint that is first order in derivatives
\begin{equation}
 \frac{2N'L'}{NL} + \frac{(N')^2}{N^2} = \alpha\left[\gamma(f')^2 + \frac{(P')^2}{2L^2} + (\gamma-2)\frac{P^2 f^2}{L^2} + \left(\gamma - \frac{3}{2}\right) \beta (f^2-1)^2\right]
\end{equation}
Finally, the modified conservation law (\ref{mod_conservation}) reads
\begin{equation}
\label{conservation}
\frac{4N'}{N} \left((f')^2+ \frac{(P')^2}{2L^2}\right) + \frac{2L'}{L}\left((f')^2-\frac{P^2 f^2}{L^2}\right) + 
\gamma \left((f')^2\right)'+\left(\frac{(P')^2}{2L^2}\right)' + (\gamma-2)\left(\frac{P^2 f^2}{L^2}\right)' 
+ \left(\gamma-\frac{3}{2}\right)\beta\left(\left(f^2-1\right)^2\right)' = 0 \  .
\end{equation}

For $\gamma=1$ we can derive the Euler-Lagrange equations by the variation of the action with respect
to the matter fields. This is not possible here, hence we ``read-off'' the equations from the 
conservation law (\ref{conservation}). This is motivated by the fact that in standard Einstein gravity
the conservation law holds {\it on shell}, i.e. for solutions of the Euler-Lagrange equations. Moreover, we require
that the equations become equal to the standard equations in the $\gamma=1$ limit and constitute a minimal extension of the
given model (see also the Appendix).

It hence makes sense to consider the following equations of motion for the matter fields:
\begin{equation}
\label{eq_higgs}
\gamma f'' + \left(\frac{2N'}{N} + \frac{L'}{L}\right) f' + \left(\gamma-2\right)
\frac{P^2 f}{L^2} + (2\gamma-3)\beta f (f^2-1) = 0 
\end{equation}
and
\begin{equation}
\label{eq_gauge}
\frac{L}{N^2}\left(\frac{N^2P'}{L}\right)' + 2(1-\gamma)\left(\frac{P^2}{P'}\frac{L'}{L} f^2\right)=2(2-\gamma) f^2 P  \  .
\end{equation}
Note that these, indeed, reduce to the standard Euler-Lagrange equations in the limit $\gamma=1$. 

Hence we have to solve a system of four coupled, non-linear ordinary differential equations numerically
subject to appropriate boundary conditions. These read
\begin{equation}
 L(0)=0 \ \ , \ \ L'(0)=1 \ \ , \ \ N(0)=1 \ \ , \ \ N'(0)=0 \ \ , \ \ f(0)=0 \ \ , \ \ P(0)=1  \  ,
\end{equation}
to ensure regularity on the $z$-axis.
Furthermore, we want the matter fields to reach their vacuum expectation values asymptotically. So, we require
\begin{equation}
 f(r\rightarrow \infty) \rightarrow 1 \ \ , \ \ P(r\rightarrow \infty)\rightarrow 0 \ .
\end{equation}

\subsection{Behaviour close to the string axis}
The behaviour at $r=0$ is determined by the requirement of imposing globally regular solutions. From the boundary 
conditions it follows for the metric functions 
\begin{equation}
 N(r) \sim 1 + n_0 r^2 \   \ {\rm and} \  \   L(r) \sim r  \  ,
\end{equation}
where $n_0$ is a constant. 

Inserting this into (\ref{eq_higgs}) we find the following behaviour of $f(r)$ at small $r$ 
\begin{equation}
\label{f_r_small}
 f(r) \sim f_0 r^{d_1}  \  \ {\rm with} \    \   
  d_1= \frac{\gamma-1}{2\gamma}\pm \sqrt{\left(\frac{\gamma-1}{2\gamma}\right)^2 - 1 + \frac{2}{\gamma}} \  .
\end{equation}
For $\gamma=1$ this reduces to $f(r)\sim f_0 r$, but here  the behaviour is much more complicated and depends
strongly on $\gamma$. In particular we note that we have to require $d_1 > 0$ in order to have
regular solutions. For the gauge field we  find
\begin{equation}
 P(r) \sim 1 + P_0 r^{2} \ \ \ .
\end{equation}
Consequently, the behaviour of the gauge field at the origin does not depend on $\gamma$. 

\subsection{Asymptotic behaviour}
In the absence of matter sources, Rastall gravity reduces to standard Einstein gravity. 
Since the energy-momentum tensor associated with the matter fields in our model falls off exponentially fast
and since we want the Abelian-Higgs string to be well-localized, we can assume that far away from the string core 
the space-time corresponds to a cylindrical
vacuum space-time. Now, it is well known that cylindrical solutions of the vacuum Einstein equations are of the Kasner type.
Hence, we would expect our solutions to asymptote to these such that
the fall-off of the metric functions is
\begin{equation}
 N(r) \sim N_1 r^a   \   \  {\rm and}  \   \      L(r) \sim L_1 r^b \  \  , 
\end{equation}
where the coefficients $a$ and $b$ have to fulfill the Kasner conditions
\begin{equation}
\label{kasner}
 2a^2 + b^2 = 2a + b = 1   \  ,
\end{equation}
with $N_1$ and $L_1$ being constants. 
The two possible solutions to (\ref{kasner}) are
\begin{equation}
(a,b)=(0,1)   \    \    {\rm and} \   \   (a,b)=\left(\frac{2}{3}, - \frac{1}{3}\right) \ \  .
\end{equation}
The first set of parameters corresponds to string-like solutions, in which case 
$L_1$ determines the deficit angle of the space-time. The second set of possible values are the so-called
Melvin solutions, which are not of physical interest in cosmological settings, however are mathematical solutions
to the equations of motion. The string-like solution then possesses a deficit angle which can be expressed as follows:
 \begin{equation}
 \label{deficit_def}
 \Delta = 2\pi \left(1-L'(\infty)\right) = 2\pi\left(1-L_1\right) \ . 
\end{equation}

The matter field functions have the following behaviour at $r\rightarrow \infty$
\begin{equation}
 f(r) \sim 1 - \frac{f_1}{\sqrt{r}} \exp(-m_{f,\gamma} r)   \   \   {\rm with}   \   \   m_{f,\gamma}=\sqrt{\beta\left(\frac{6}{\gamma}-4\right)}
\end{equation}
and
\begin{equation}
P(r) \sim P_1 \sqrt{r} \exp\left(-m_{P,\gamma} r\right)  \   \  {\rm with} \   \    m_{P,\gamma}=\sqrt{4-2\gamma}  \   , 
\end{equation}
where $f_1$ and $P_1$ are constants and $m_{f,\gamma}$ and $m_{P,\gamma}$ correspond
to the effective Higgs and gauge boson mass, respectively.
For $\gamma=1$ the Higgs field reaches its vacuum value quicker than the gauge field if $\beta > 1$ and slower if $\beta < 1$.
The value $\beta=1$ corresponds to the BPS limit. In this limit the masses of the  gauge and Higgs bosons are equal. 
Now for $\gamma < 1$ this changes.
The effective Higgs and gauge boson mass are equal for $\beta_{\rm equal}=(2\gamma-\gamma^2)/(3-2\gamma)$, which
is equal unity for $\gamma=1$ and decreases monotonically with decreasing $\gamma$.

\section{Numerical results}
\label{Numerical}

To the best of our knowledge there are no explicit solutions to the set of coupled differential equations presented above.
We have hence solved the equations numerically using the ODE solver COLSYS \cite{colsys}.
Relative errors of the solutions are typically on the order of $10^{-8}$ to $10^{-10}$ (and sometimes even better).

The limit $\gamma=1$ corresponds to standard Einstein gravity. From (\ref{eq_higgs}) it is apparent
that $\gamma=0$ is excluded. Furthermore, we see from (\ref{f_r_small}) that in order for the
solutions at $\gamma=1\pm \delta$, where $\delta$ is small,  to behave like the solutions
in the $\gamma=1$ limit we must choose the positive sign. This means that for $\gamma > 1$ the 
parameter $ 0 < d_1 < 1$. However, since we need to require $L(r) \sim r$ at $r\sim 0$, otherwise
the space-time would not be regular,  this leads to infinities on the right hand side of the gravity
equations (\ref{eq_grav1}) and (\ref{eq_grav2}). So, we conclude that we have to choose $\gamma \in (0, \  1]$.

For $\gamma=1$ it is well known that the 
coupled system of equations admits gravitating Abelian-Higgs string solutions. In this paper we are interested to investigate the influence
of the Rastall parameter, $\gamma$, on the behavior of the matter fields and on the metric. 
Since Rastall gravity reduces to Einstein gravity in the absence of sources
our asymptotic space-time (in which the matter fields reach their vacuum values) should be a solution to Einstein gravity.
We can hence use the standard definition of a deficit angle given in (\ref{deficit_def}). 

Let us first recall what is know about the Einstein gravity case $\gamma=1$. 
In this case, it has been observed that the value of $\Delta$ depends on  both $\beta$ and $\alpha$ and increases
with increasing $\alpha$. At some maximal value of $\alpha=\alpha_{\rm max}$ the deficit angle
becomes equal to $2\pi$. For $\alpha > \alpha_{\rm max}$ no globally regular
string solutions exist, but only solutions with singularities (so-called ``supermassive" or ``inverted" string solutions).
These supermassive solutions possess a singularity at some finite value of the radial coordinate
$r=r_{\rm max,1}$ at which $L(r_{\rm max,1})=0$, while
$N(r_{\rm max,1})$ stays finite \cite{clv,Brihaye:2000qr}.

\begin{figure}[h]
\begin{center}

\subfigure[][gauge field function $P(r)$]{\label{p_gamma}
\includegraphics[width=8cm]{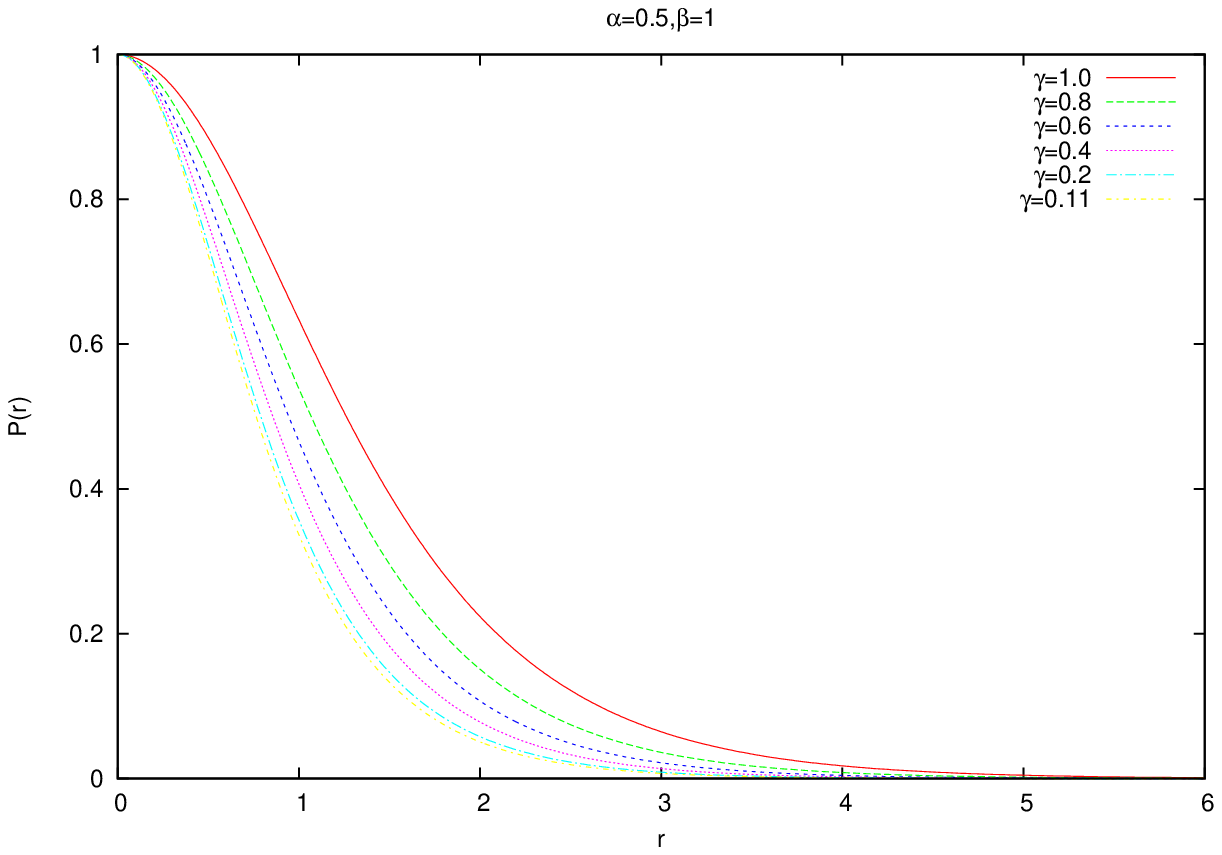}}
\subfigure[][Higgs field function $f(r)$]{\label{f_gamma}
\includegraphics[width=8cm]{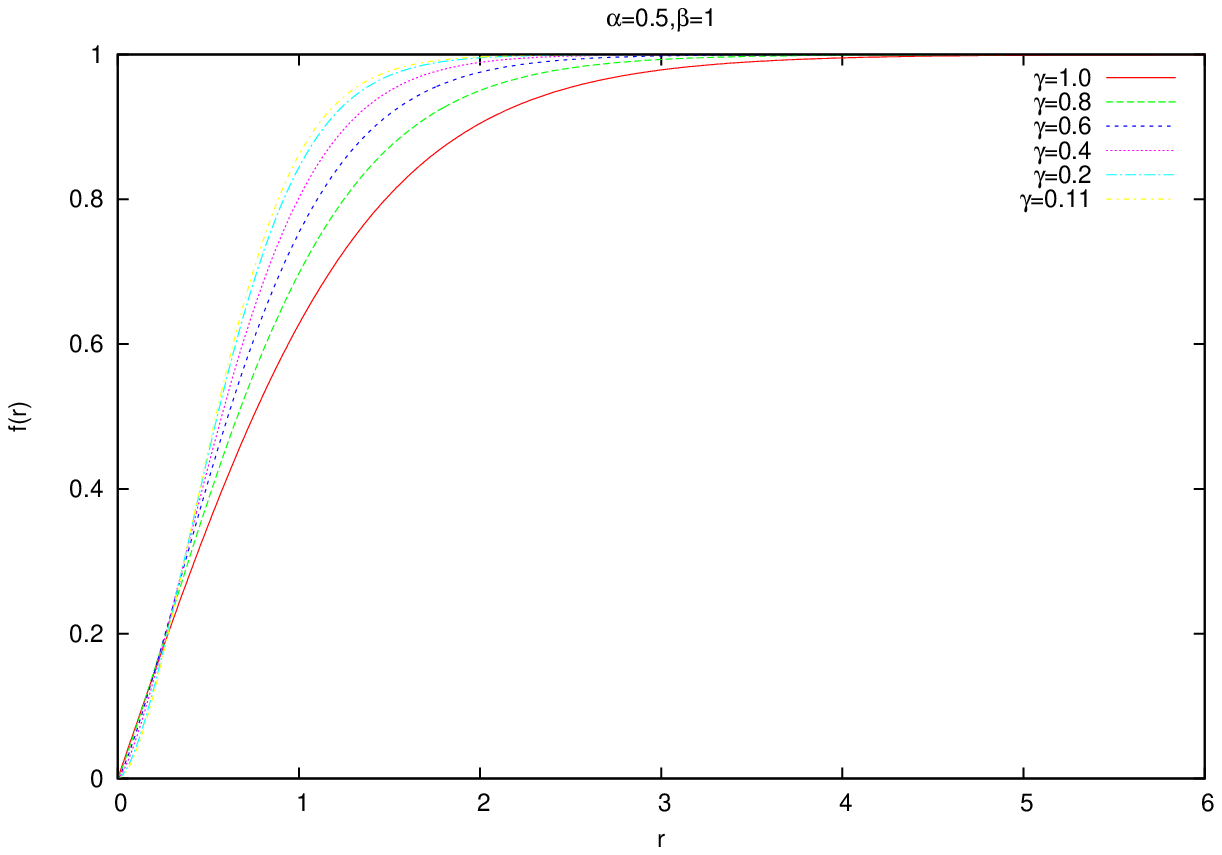}} \\
\subfigure[][metric function $N(r)$]{\label{n_gamma}
\includegraphics[width=8cm]{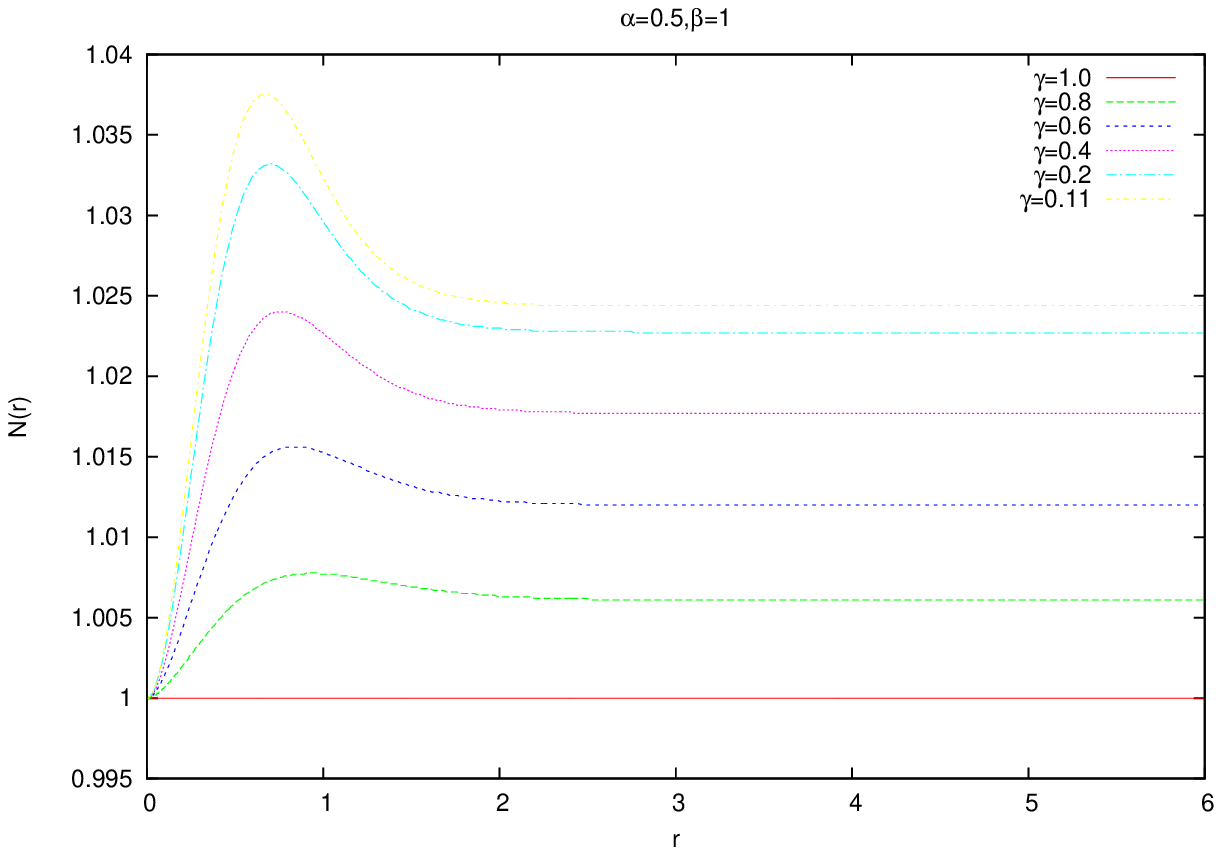}}
\subfigure[][metric function $L(r)$]{\label{l_gamma}
\includegraphics[width=8cm]{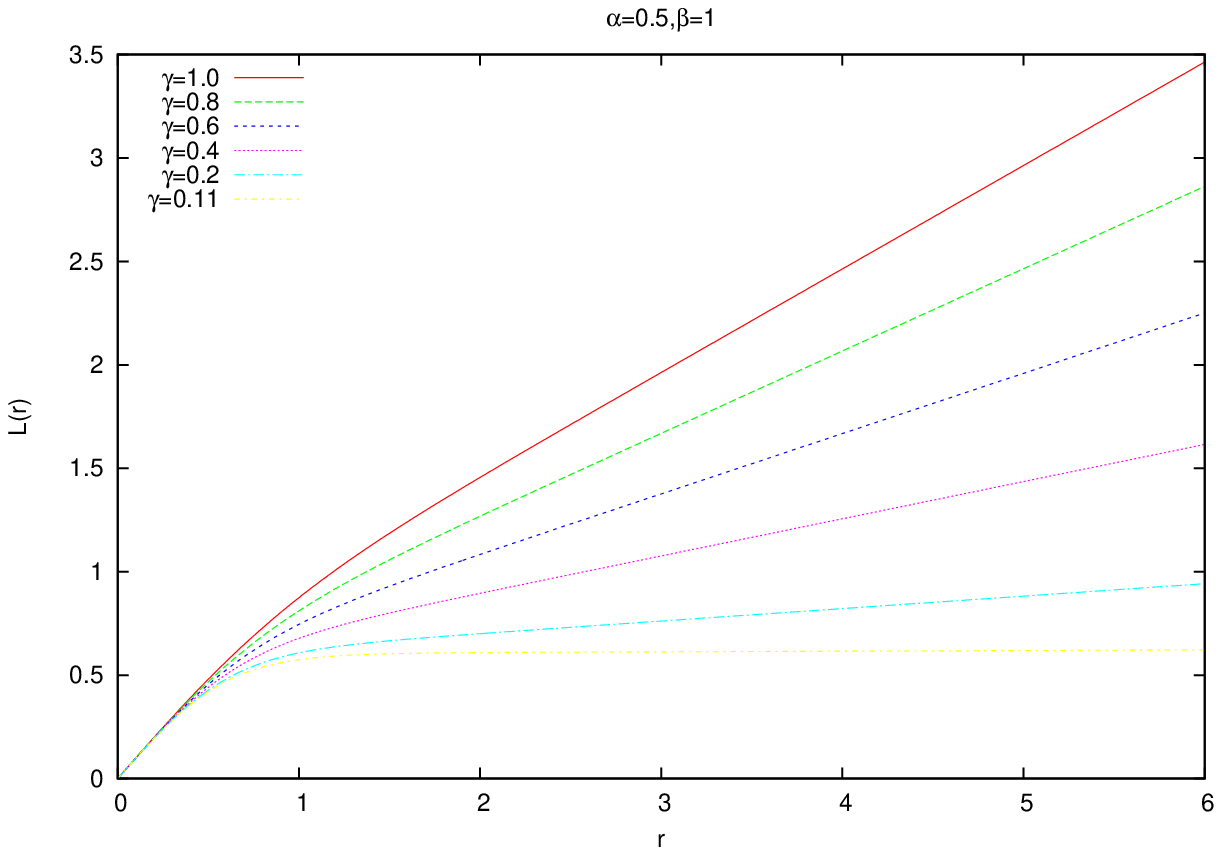}}
\end{center}
\caption{\label{profiles} We show the profiles of the matter functions $P(r)$ and $f(r)$ (top) and the metric
functions $N(r)$ and $L(r)$ (bottom) for Abelian-Higgs strings in Rastall gravity with $\alpha=0.5$ and $\beta=1.0$. The
$\gamma=1$ curves correspond to the Einstein gravity limit, while the $\gamma=0.11$ case gives the profiles of the 
solution which has $\Delta\approx 2\pi$. }    
\end{figure}

In Fig.\ref{profiles} we show the behavior of a typical Abelian-Higgs string solution for $\beta=1$ and $\alpha=0.5$ and different
values of $\gamma$. For $\beta=1$ and $\gamma=1$ these solutions fulfill a Bogomolnyi-Prasad-Sommerfield
(BPS) bound. In this limit the components of the energy-momentum tensor in the direction perpendicular to the string
axis vanish implying via (\ref{eq_grav1}) that $N(r)\equiv 1$. This means that there is no gravitational force acting perpendicular to
the string axis. In this limit, the remaining equations can be recast into the form
\begin{equation}
 f'= \frac{P}{L} f \ \ , \ \  P'=L(f^2-1) \  \ {\rm and} \ \ 
L''=-\alpha L \left((f^2-1)^2 + \frac{2P^2 f^2}{L^2}\right)  \ , \ \ \ {\rm for} \ \ \gamma=\beta=1  \  .
\label{bps}
\end{equation}
A solution of this type is shown in Fig.\ref{profiles}. Decreasing the Rastall parameter $\gamma$ we observe that both the 
gauge field function $P(r)$ as well as the Higgs field function $f(r)$ are stronger localized around the string axis implying that
the width of the string decreases. At the same time, the metric function $N(r)$ starts to deviate from its constant
value of unity stronger and stronger when decreasing $\gamma$. Furthermore, the metric function $L(r)$ possesses a decreasing slope at large $r$
implying an increase in the deficit angle $\Delta$ with decreasing $\gamma$. This seems natural since the energy-momentum content 
is localized inside a smaller region of space-time.

Let us now discuss the value of the deficit angle, $\Delta$,  in more detail since this is an important quantity
when predicting observational consequences of strings. The deficit angle leads to gravitational lensing
as well as to red-and blue-shift of photons towards which and away from which, respectively, the string is moving (the Kaiser-Stebbins
effect). Strings (if they existed) would hence have an important impact on the temperature anisotropies of the CMB.
It is often stated that the deficit angle $\Delta = 8\pi G \tilde{\mu}= \alpha\mu$. This is strictly speaking only
true in the BPS limit $\beta=1$. In this case, it is easy to see from (\ref{bps}) and the definition of $\mu$ that 
this relation indeed holds. Since furthermore in this limit, the energy per unit length in the BPS limit is $\mu=2\pi$ we find that the deficit
angle in this specific case is $\Delta=2\pi \alpha$.

\begin{figure*}[h!]
\begin{center}
\includegraphics[width=12cm]{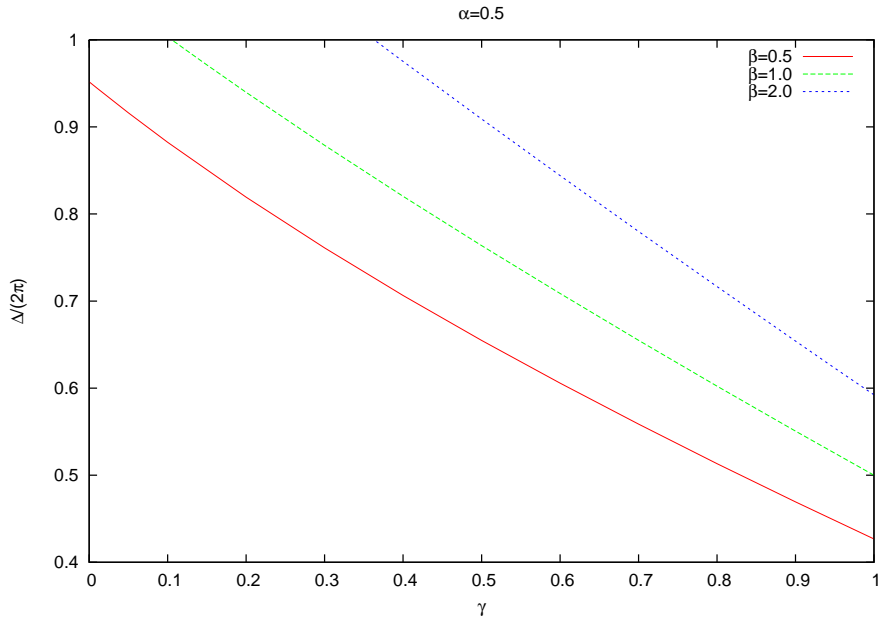}
\end{center}
\caption{We show the value of the deficit angle $\Delta=2\pi\left(1-L'(\infty)\right)$
in dependence on $\gamma$ for $\alpha=0.5$ and different
values of $\beta$. \label{alpha05_string}}
\end{figure*}

We have studied the case $\alpha=0.5$ and $\beta=0.5$, $\beta=1$ and $\beta=2$, respectively. Our results
are shown in Fig.\ref{alpha05_string}, where we give the deficit angle $\Delta$ as function of $\gamma$. 
The function $N(r)$ stays finite all along and varies only little. This is why we do not present any detailed results about it here. 

For $\gamma=1$ the deficit angle has the known value (see e.g.\cite{Brihaye:2000qr}).
E.g. in the BPS limit, $\beta=1$,  we have $\Delta = \pi$. Decreasing $\gamma$ we find that
the deficit increases until it reaches $\Delta=2\pi$ at some value of $\gamma=\gamma_{\rm cr}$.
For $\gamma < \gamma_{cr}$ the solutions have a singularity at a finite value of the radial
coordinate $r=r_{\rm max}$ with $L(r=r_{\rm max})=0$. 

\begin{figure*}[h!]
\begin{center}
\includegraphics[width=12cm]{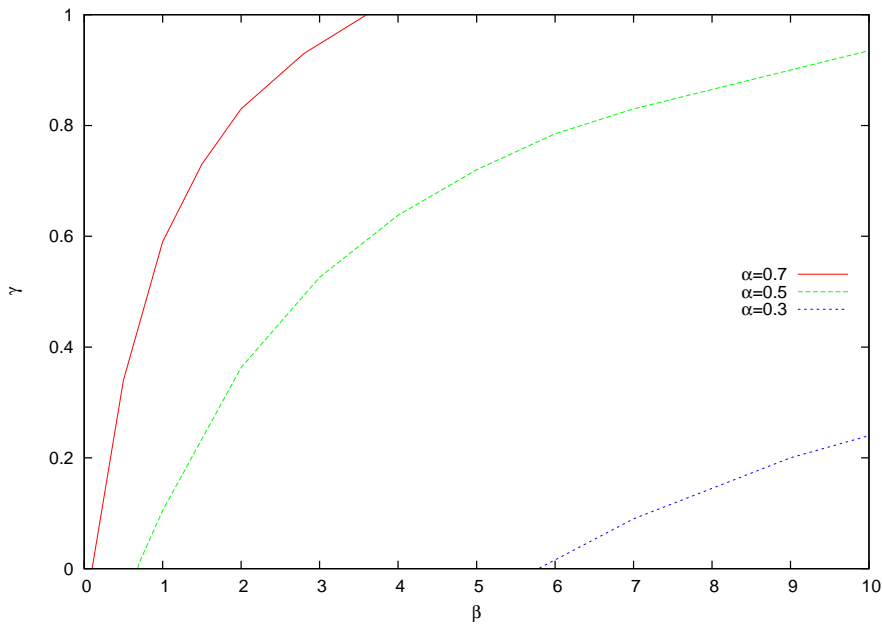}
\end{center}
\caption{We show the value of $\gamma_{cr}$ in dependence on $\beta$ for three different values of $\alpha$. 
String solutions with deficit angle smaller than $2\pi$ and hence without singularity exist only for values of
$\gamma$ above the curve. \label{fig_gammacr}}
\end{figure*}

The critical value for $\gamma$ depends on the value of $\beta$. Considering $\alpha=0.5$, we observe that 
for $\beta=2$ the critical value is $\gamma_{\rm cr}\approx 0.363$, while
for $\beta=1$ we have $\gamma_{\rm cr}\approx 0.105$. For $\beta=0.5$ we find that the
deficit angle $\Delta$ stays smaller than $2\pi$ for all values of $\gamma\in (0, \ 1]$. The results
for $\alpha=0.5$ are shown in Fig.\ref{fig_gammacr}. For $\beta \lesssim 0.7$ we find that non-singular string
solutions exist for all values of $\gamma\in (0, \ 1]$. For increasing $\beta$ the value of $\gamma_{cr}$ increases, but reaches $\gamma=1$ 
only exponentially slow. Hence, we find that for all reasonable values of the Higgs to gauge boson mass ratio
as well as the ratio between the symmetry breaking scale and the Planck mass {\it regular} Abelian-Higgs strings in Rastall gravity can
be constructed.

\begin{figure*}[h!]
\begin{center}
\includegraphics[width=12cm]{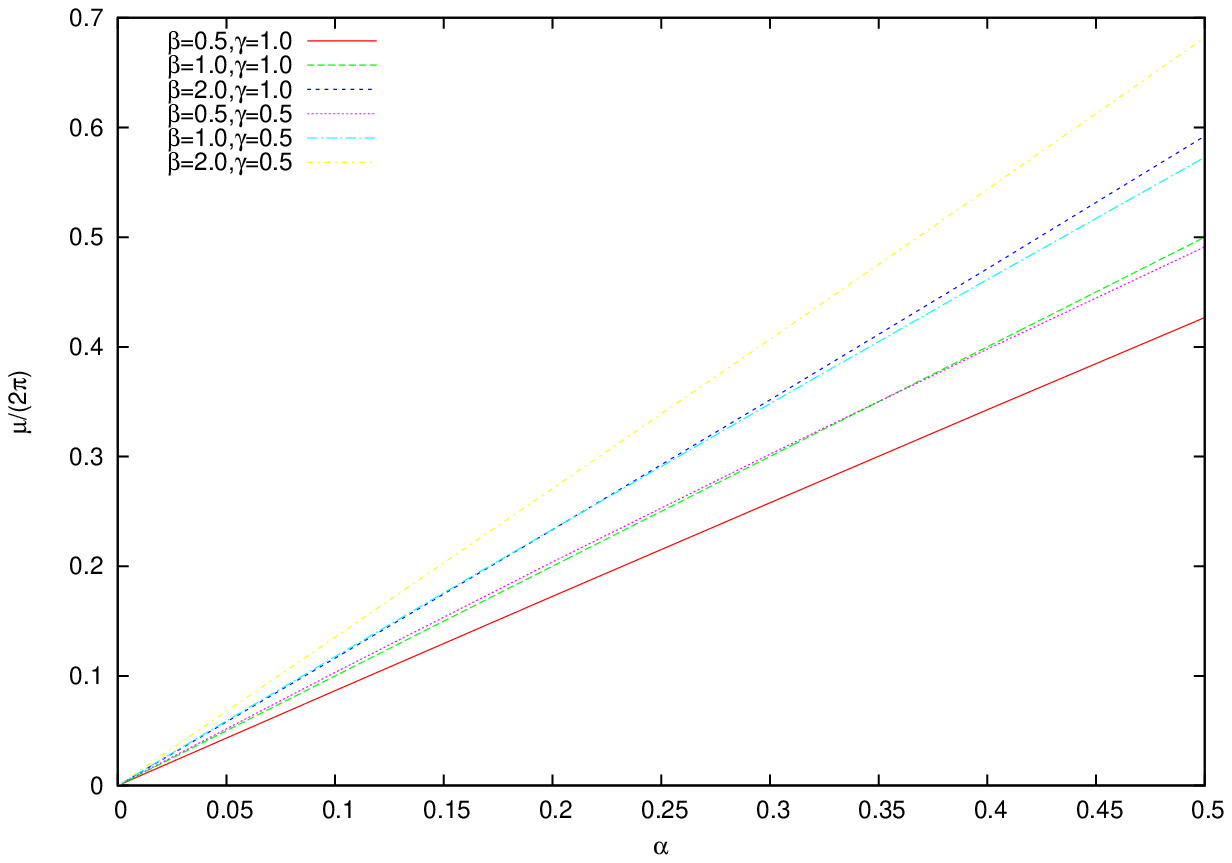}
\end{center}
\caption{We show the value of the energy per unit length $\mu/(2\pi)$ in dependence on $\alpha$ for three different values of $\beta$
and two different values of $\gamma$, respectively. \label{fig_mu}}
\end{figure*}

As stated above, for $\gamma\neq 1$ and/or $\beta\neq 1$ there is no linear relation between the
deficit angle $\Delta$ and the energy per unit length $\mu$. We have hence also studied the energy per unit length in
dependence on the parameters of the model. Our results are shown in Fig.\ref{fig_mu}.

As is clearly seen from this figure, we have $\mu=2\pi\alpha$ for $\gamma=\beta=1$. For the other cases,
the energy per unit length still depends (nearly) linearly on $\alpha$ and behaves as 
\begin{equation}
 \mu=2\pi \left(1+\epsilon(\beta,\gamma)\right)\alpha \  ,
 \label{energy-density}
\end{equation}
where $\epsilon(\beta,\gamma)$ is a function that depends on $\beta$ and $\gamma$ and fulfills the condition $\epsilon(1,1)=0$.
Now, we find that this function is monotonically increasing with $\beta$ and becomes nearly constant for very large
values of $\beta$. For $\beta < 1$ it is negative, while for $\beta > 1$ it is positive in the Einstein gravity limit $\gamma=1$. 
This function also increases with decreasing $\gamma$, i.e. moving away more and more from the Einstein gravity
limit. The increase is stronger for larger values of $\beta$. Now, we can make an interesting observation.
If we choose $\beta < 1$,  we can find values of $\gamma$ for which $\mu=2\pi$, i.e. corresponding to the energy per unit length
that the solution has in the BPS limit in standard Einstein gravity. In the following we will refer to this as a {\it would-be-BPS limit}.
For $\gamma=\beta=1$ this is true for all values of $\alpha$. For a fixed value of $\beta < 1$ and $\alpha$ we can then decrease
$\gamma$ such that at some specific value of $\gamma=\tilde\gamma$,  the energy per unit length becomes again equal to unity (in units of
$2\pi$). We find that the lower $\beta$ the lower we have to choose $\gamma$ to achieve this condition. E.g. for $\alpha=0.1$
we find that $\tilde\gamma\approx 0.73$ for $\beta=0.75$, while $\tilde\gamma\approx 0.56$ for $\beta=0.5$. We 
also find a (albeit smaller) dependence on $\alpha$. E.g. for $\beta=0.5$ we find $\tilde\gamma\approx 0.45$ for $\alpha=0.5$.
Hence, $\tilde\gamma$ decreases with increasing $\alpha$.

\section{Conclusions}
\label{Conc}

In this paper we have studied Abelian-Higgs strings in the context of Rastall gravity.
Rastall theory touches one of the cornerstones of General Relativity, namely the 
conservation of the energy-momentum tensor.
The violation of the conservation law is parametrized in terms of a parameter $\gamma$ with $\gamma=1$ constituting the
General Relativity limit. 
In spite of its phenomenological character Rastall gravity may be related to an effective (and hence classical) 
implementation of a gravitational anomaly that might appear due to quantum effects. 
Our main purpose here was to investigate the impact of these effects on a field theoretical realization of line-like topological
defects, so-called cosmic strings. Our main results can be summarized as follows: 
\begin{itemize}
 \item we find that singularity-free space-times are possible only when $0 < \gamma \leq 1$,
 \item depending on the other parameters in the model (the two ratios between the fundamental mass scales)
       the solid deficit angle becomes equal to $2\pi$ at a value of $\gamma_{\rm cr} > 0$,
       \item a BPS limit in which the energy per unit length saturates a bound (and hence becomes equal to $2\pi n$) 
       does not seem to exist here unlike for the Einstein gravity 
       limit, where
       it exists for equal gauge and Higgs boson mass,
       \item a {\it would-be-BPS bound} exists at which the energy per unit length becomes equal to $2\pi n$,
       i.e. fulfills the above mentioned bound. However, this
       is not related to an underlying mathematical structure.
\end{itemize}

Our results are interesting because of the recently presented BICEP2 data. If the measurements of the B-mode polarization are
confirmed (preferably additionally through other measurements like e.g. the PLANCK collaboration) we do have a window to the
very early universe and the phase of inflation. Now inflation seems to be driven by scalar fields and the question remains
where these originate from. Most unifying models that are able to model inflation predict the production of cosmic strings
at the end of inflation. Hence, it might turn out that if inflation took place, cosmic strings become a prediction rather than a speculation.
Since it is certain that the energy conditions at the epoch of inflation are extreme, we would expect that quantum effects
play a r\^ole (even if one treats the gravity side classically this can certainly not be said for the matter side). 
Rastall gravity is one possibility to model these quantum effects effectively. Our results presented above suggest that 
taking such corrections into account could have stronger effects on the CMB due to an increased deficit angle. 

While in this paper we have studied string-like objects without additional structure one could also think of investigating
strings with additional degrees of freedom inside their core. These could be in the form of fermionic or bosonic currents and the
corresponding objects have been coined {\it superconducting strings} \cite{witten2}. 
Since the stability of these objects is of huge importance in the context of the formation of loops of cosmic string, so-called {\it vortons}
(see e.g. \cite{peter_uzan} for more details)
a macroscopic stability criterion has been developed \cite{carter,carter2} and used in a detailed
analysis for superconducting string solutions
of the U(1)$\times$ U(1) model in flat space-time \cite{patrick3} as well as in curved space-time \cite{hartmann_michel}. 
It would be very interesting to study possible quantum effects on the stability of these objects and the Rastall gravity model
would implement these quantum effects naturally. 
\clearpage
{\bf Acknowledgment}  B.H. would like to acknowledge the 
CNPq for financial support. B.H. would also like to acknowledge 
the Deutsche Forschungsgemeinschaft (DFG) for
support within the framework of the DFG Research Group 1620 {\it Models of Gravity}.
E.R.B.M. and J.C.F. would also like to acknowledge 
CNPq and FAPES for partial financial support.

\clearpage

\section{Appendix: the conservation law}
In order to write down (\ref{mod_conservation}) explicitly  we need the non-vanishing Christoffel symbols associated
to the metric and in particular - since the energy-momentum tensor is diagonal - only the Christoffel symbols
with equal lower indices. The non-vanishing Christoffel symbols with equal lower indices are
\begin{equation}
\label{christoffel}
\Gamma_{tt}^r = N'\cdot N \ \ , \ \ \Gamma_{\varphi\varphi}^r = - L'\cdot L \ \ \ , \ \ \ \Gamma_{zz}^r = -N\cdot N' \ .
\end{equation}
The conservation law (\ref{mod_conservation}) reads explicitly taking only the non-vanishing terms into account
\begin{equation}
 \frac{1}{\sqrt{-g}}\partial_r \left(\sqrt{-g} T^{rr}\right) + \Gamma_{\lambda\lambda}^r T^{\lambda\lambda} = 
\frac{1-\gamma}{2} \partial_r T  \ .
\end{equation}
Now using that $\sqrt{-g}=N^2L$ and (\ref{christoffel}) we find 
\begin{equation}
 N^2 L (T^{rr})' + \left(2N N'L + N^2 L' \right) T^{rr} + N'N^3 L T^{tt} - L'N^2 L^2 T^{\varphi\varphi}  
- N'N^3 L T^{zz} = \frac{1-\gamma}{2} N^2 L T'
\end{equation}
where the prime denotes the derivative with respect to $r$. 
  This can be rewritten in the following form
  \begin{eqnarray}
  & & 2f'\left(\gamma f'' + \frac{L'}{L} f' + \frac{2N'}{N} f' + (\gamma-2)\frac{f P^2}{L^2} + (2\gamma-3) \beta f (f^2-1)\right) \nonumber \\
  &+& P'\left(\frac{2 N'}{N} \frac{P'}{L^2} + \frac{P''}{L^2} - \frac{L'P'}{L^3} + 2(\gamma-2)\frac{Pf^2}{L^2}
  - 2(\gamma-1)\frac{L'}{L^3} \frac{P^2 f^2}{P'}\right) = 0  \ ,
   \end{eqnarray}
  where we have sorted terms with respect to pre-factors of $f''$ and $P''$, which are $f'$ and $P'$ respectively. This then leads to
  quasi-linear 2nd order differential equations in the fields $f$ and $P$. 
  Now, we require the respective terms in the brackets to vanish. This gives the equations (\ref{eq_higgs}) and (\ref{eq_gauge}).
  We believe that this is a suitable choice for our model. We insist that the model we are using here is phenomenological and as such
  our choice of equations is one possible choice amongst others. 
  The use of phenomenological models to describe quantum effects in curved space-time is a widely used procedure since - up to now -
  a fully consistent model of quantum gravity does not exist.
\end{document}